# Crowd Science: Measurements, Models, and Methods

John Prpić
Lulea University of Technology
john.prpic@ltu.se

Prashant Shukla
Accenture Institute for High Performance, Boston
prashant.p.shukla@accenture.com

## Abstract

*The increasing practice of engaging crowds, where organizations use IT to connect with dispersed individuals for explicit resource creation purposes, has precipitated the need to measure the precise processes and benefits of these activities over myriad different implementations. In this work, we seek to address these salient and non-trivial considerations by laying a foundation of theory, measures, and research methods that allow us to test crowd-engagement efficacy across organizations, industries, technologies, and geographies. To do so, we anchor ourselves in the Theory of Crowd Capital, a generalizable framework for studying IT-mediated crowd-engagement phenomena, and put forth an empirical apparatus of testable measures and generalizable methods to begin to unify the field of crowd science.*

## 1. Introduction

A multitude of endeavors from our modern world exemplify the usefulness of engaging crowds through IT. Wikipedia, reCAPTCHA, Yelp, Quirky, Uber, Threadless, Waze, AirBnb, and Trip Advisor, among many others, are frequently cited as examples of the power and utility of engaging IT-mediated crowds [1, 2, 3]. Similarly, numerous intermediaries providing access to IT-mediated crowds as a service, such as eYeka, Kickstarter, TopCoder, M-Turk, Crowdflower, Innocentive, and Tongal, have also been praised for their efficacy in achieving specific solutions for their clients [4,5,6,7].

Concomitantly, in research we have seen the emergence of disparate streams of literature, across numerous disciplines, concerned with investigating one or more aspects of IT-mediated crowd engagement in both the private and public sectors. Crowdsourcing [3,4], Citizen Science [8], Prediction Markets [9], Open Innovation platforms [7], Crowdfunding [10], Peer Production [11] and Human Computation [12,13] all illustrate emerging areas of research where IT-mediated crowds are creating new socio-technical configurations, potentials, and outcomes.

However, research that bounds all the extant research silos with all the examples of the organizational use of IT-mediated crowds in public and private sector practice, has yet to emerge. As an effort to solve this salient research gap, this work introduces the details of an empirical and generalizable crowd science research program that can unite all the disparate branches of research and practice on IT-mediated crowds. To do so, we introduce the details of an empirical apparatus comprised of models, measurements, and research methodologies suitable for a unified crowd science.

In the ensuing sections of this work, we achieve these goals by first introducing the theoretical background upon which a crowd science can be built, outlining in detail the Theory of Crowd Capital [1,2] as our current governing model of crowd science. In section 3, we detail the challenges in operationalizing, generalizing, and testing the benefits of IT-mediated crowd-engagement through the Theory of Crowd Capital (TCC). In section 4, we put forth potential operationalizations for each of the constructs and dimensions of the TCC to illustrate specific measures that can be used to test crowd-engagement processes and outcomes. In section 5, we detail different research methods that can be implemented to advance crowd science toward causally asserting the benefits of IT-mediated crowd-engagement. In section 6, we discuss the limitations of the crowd science paradigm introduced, before concluding with a summary of the contributions of our research.

## 2. Crowd Science Background

There is a significant amount of research emerging in disparate fields on the organizational use of IT-mediated crowds for resource creation purposes. Crowdsourcing [3, 4, 14, 15, 16], Citizen Science [8], Prediction Markets [9], Open Innovation platforms [7, 17], Crowdfunding [10], Peer Production [11] and Human Computation [12, 13], all illustrate emerging areas of research where IT-mediated crowds are studied.

Although it is beyond the scope of this work to give a rich and detailed account of the similarities and differences of each of these fields, as others have already begun to do [1, 64], we'll briefly outline the evolution of the literature in the field of Crowdsourcing in particular, since it is perhaps the richest and broadest extant body of literature on IT-mediated crowds, and numerous efforts in this particular field have already attempted to integrate the various research silos listed above [16, 18, 19, 20, 21, 64].

**2.1 Crowdsourcing**

Brabham first described Crowdsourcing as the organizational use of IT to engage crowds comprised of groups and individuals, for the purposes of completing tasks, solving problems or generating ideas. Wherein Crowdsourcing is a deliberate blend of bottom-up crowd-derived processes and inputs, combined with top-down goals set and initiated by an organization [3,4]. As an IT-mediated problem solving, idea-generation, and production model for organizations, Crowdsourcing leverages the distributed knowledge found in crowds [3,22], through different means such as micro-tasking [12, 13], open collaboration [16] or tournament-based competitions [23, 24].

From this research beginning, the body of Crowdsourcing research has expanded rapidly, supplying numerous taxonomies, typologies, and findings related to the various processes, attributes, and outcomes of engaging IT-mediated crowds, including some of the following themes; task complexity [25], crowdsourcing models [19], crowdsourcing processes [18], crowd ability [26], solution quality [27], crowdsourced data [28] and data processing [29], enterprise crowdsourcing [30], crowdsourcing as a lens for human-computer interaction [31], organizational crowdsourcing intentions [23, 32], value creation [33], crowdsourcing multiple tasks [34], crowdsourcing for behavioral science purposes [35], crowdsourcing and algorithms [36], crowdsourcing for innovation [37], crowdsourcing labor law [38], crowdsourcing workers with disabilities [39], health care crowdsourcing [38], crowdsourcing for policy assessment [40], the geography of crowdsourcing participation [62], and cultural factors in crowdsourcing [41].

Most recently, the Crowdsourcing research is beginning to coalesce, organizing the almost bewildering array of Crowdsourcing research emerging, into three types of generalized Crowdsourcing applications available for organizational use; virtual labor markets, tournament-crowdsourcing, and open collaboration. Further, each of the three generalized forms of Crowdsourcing application can be compared along seven universal characteristics (see list below adapted from [42]):

1) Cost of using a Crowdsourcing technique
2) Anonymity of Crowdsourcing participants
3) Scale of crowd size
4) IT Structure of Crowdsourcing application
5) Time required to implement Crowdsourcing
6) Magnitude of Crowdsourcing tasks
7) Reliability of the Crowdsourcing technique

The universal characteristics of all Crowdsourcing applications listed above are useful and important, since they allow researchers and practitioners alike to understand the stable, relative differences between the forms of Crowdsourcing, while vividly displaying the inherent trade-offs that organizations face when considering Crowdsourcing initiatives. Further, recent work (see Figure 1 below borrowed from [16]), has similarly clarified the vast variety of Crowdsourcing research in respect to the types and characteristics of the work that can be achieved by organizations through IT-mediated crowds [16].

**Figure 1 – Typology of Crowdsourcing Work**

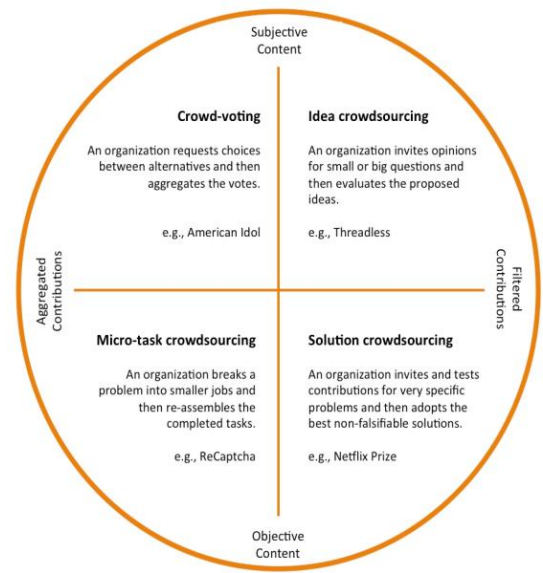

Simultaneous to these advances in the Crowdsourcing literature, other work has emerged that attempts to subsume all IT-mediated crowd phenomena into a parsimonious framework. In the ensuing section of this work we introduce the Theory of Crowd Capital as an established and validated framework from the literature, which allows researchers to investigate all implementations of IT-mediated crowds for organizational resource creation.

## 2.2 Crowd Capital

The resource based view (RBV) [43] and the knowledge-based view of organizations (KBV) [44, 45], assert that rare and inimitable knowledge is a valuable resource, potentially giving organizations an advantage over their competitors. Embedded in the theoretical context of the RBV and the KBV is the Theory of Crowd Capital (TCC) [1, 2, 16, 42], which bounds and explains the dynamics and mechanisms that enable organizations to engage crowds through IT for resource creation purposes (see Figure 2 below, adapted from [63], and Figure 3 below from [16]).

**Figure 2 – The Theory of Crowd Capital – Constructs**

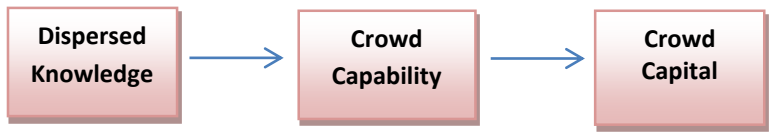

As a tool of managerial or research use, the TCC framework is multi-purpose one, due to its heritage in the innovation, economics, strategy, IS and HCI literatures. From the innovation literature TCC includes the major elements of IT-mediated absorptive capacity [16, 44]; the organizational acquisition and assimilation of knowledge. From the strategy and economics literatures TCC's antecedent condition; dispersed knowledge, stems from Hayek [46], and the overall capability and knowledge resource perspective of the framework from [43]. Further, the IS and HCI literatures are integrated into the TCC in the structure dimension of the Crowd Capability construct, as detailed in section 3.

Given the specificity of the TCC framework, and its wide applicability to IT-mediated crowd research, in this work, we choose the Theory of Crowd Capital as the first useful edifice upon which a distinct crowd science can be built. As far as we are aware, the TCC is the only organizational-level model in the research that is generalizable and falsifiable for all situations of organizations engaging IT-mediated crowds for resource creation purposes. And given that initial validations [42, 47] illustrate TCC's usefulness in guiding fine-grained empirical inquiry, we feel well supported in choosing it as the first useful model upon which to build a distinct crowd science.

**Figure 3 – The Theory of Crowd Capital – Systemic View**

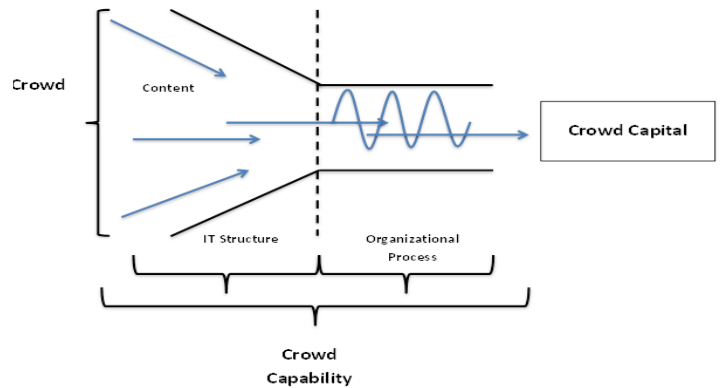

In the next section of this work, we will use this theoretical platform to outline the challenges in operationalizing, generalizing, and testing the benefits of IT-mediated crowd-engagement.

## 3. Constructs of a Crowd Science Model

Given the broad range of crowd-engaging applications that organizations can utilize, a theoretical model that generalizes the processes and dynamics of all of these specific instantiations is essential to develop a science of IT-mediated crowds, as only such a model would incorporate testability and predictability [48, 49]. Staying close to the objective of achieving these research ideals, we will next consider each of the constructs of the TCC model, and elucidate the challenges and opportunities that they bring forth as a basis for crowd science.

### 3.1 Crowds

TCC takes as given that dispersed knowledge [46] is the antecedent condition of Crowd Capital generation for organizations. For all practical purposes, dispersed knowledge is the state of nature in society, where all individuals possess some unique, private knowledge relative to all others. F.A. Hayek, describes dispersed knowledge as "…the knowledge of the circumstances of which we must make use never exists in concentrated or integrated form but solely as the dispersed bits of incomplete and frequently contradictory knowledge which all the separate individuals possess" [46].

Dispersed knowledge in its multifarious guises is the *raison d'etre* for crowd science, and because dispersed knowledge changes from moment-to-moment in every individual, the knowledge contained in any particular crowd is likewise never static either.

This is powerful! But, also problematic for researchers and practitioners.

Therefore, the initial challenges of a crowd science are quantifying and qualitatively differentiating one IT-mediated crowd from another. Is there an optimum size of a crowd? How do I construct a crowd? How do I maintain a crowd? These are some of the outstanding questions facing practitioners and crowd science researchers.

### 3.2 Crowd Capability

The TCC introduces Crowd Capability as an organizational-level construct comprised of the three dimensions discussed in turn below.

#### 3.2.1 Content

The content dimension of the Crowd Capability construct represents the form of data, information, or knowledge that an organization seeks from a crowd. Well-known forms of content that are currently being sought from crowds include micro-tasks [51], ideas and creativity [41], money [10] and innovative technical solutions [24].

#### 3.2.2 IT-Structure

The IT-structure dimension of the Crowd Capability construct indicates the technological means employed by an organization to engage a crowd. And crucially, IT-structure can be found to exist in either Episodic or Collaborative form, depending on the interface of the IT used to engage a crowd.

With Episodic IT-structures, the individual members of a crowd population never interact with each other directly through the IT. A prime example of this type of IT-structure is Google's reCAPTCHA [51], where Google accumulates significant resources from a crowd of millions of people, though it does so, without any need for the people in their crowd to interact directly with one another through the IT.

On the other hand, Collaborative IT-structures require that crowd members interact with one another through the IT, for organizational resources to be generated. Therefore, in Collaborative IT-structures, social capital must exist (or be created) through the IT for resources to be generated by the organization. A prime example of this type of crowd IT-structure is Wikipedia (or more generally wikis), where the crowd members build directly upon each other's contributions through time.

#### 3.2.3 Processing

The process dimension of the Crowd Capability construct refers to the internal procedures that an organization will use to organize, filter, and integrate the incoming crowd-derived contributions. "Successfully engaging a crowd, and successfully acquiring the desired contributions from a crowd, are necessary, but are not sufficient alone to generate crowd capital" [16]. The last mile in Crowd Capital creation is the processing of crowd contributions by an organization, and it is the process dimension of the Crowd Capability construct that is the sufficient condition of the TCC model.

### 3.3 Crowd Capital

The Crowd Capital construct is a heterogeneous organizational-level resource generated from IT-mediated crowds. It is derived from dispersed knowledge (embodied in the people in a crowd) and is a key resource (a form of capital) for an organization that can facilitate productive and economic activity [1].

Crowd Capital is a potential outcome of IT-mediated crowd engagement, and like the other forms of capital in the literature, (social capital, financial capital, human capital etc.), Crowd Capital requires investment (for example in Crowd Capability dimensions), and potentially leads to literal or figurative dividends for the organization. Crowd Capital is the central benefit that organizations seek when they engage with a crowd.

Just as there is variety in the antecedents—Dispersed Knowledge and Crowd Capability—there is considerable diversity in the type of Crowd Capital that organizations seek and create. This is expected: different organizations across industries and geographic locations seek different resources from crowds. Furthermore, the same organization might have different resource needs at different points in its lifecycle. How then do we operationalize Crowd Capital for a sound crowd science?

## 4. Operationalizing Crowd Science

In this section we tackle the first step necessary toward the development of crowd science; operationalizations (see Figure 4 below), and we outline the measurements in a step-by-step manner.

### 4.1 Measuring a Crowd

As we have already illustrated, the dispersed knowledge of every individual is unique at every given point in time, due to the spatial-temporal nature of human life as we know it. To manage this large variation, we suggest dividing a crowd into attributes that we can control for in our models.

**Figure 4 – Examples of Operationalizations for Crowd Science**

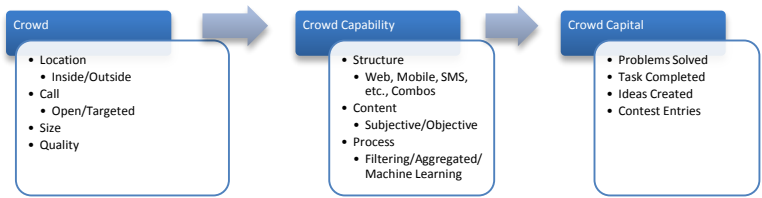

First, it may be possible to control for whether the crowd being engaged exists inside or outside of the organization. All else equal, such dummy variables can be assumed to be the treatment variable in the model, and the efficacy of external vs. internal crowds can thus be tested.

Second, the call, or the appeal made by an organization to individuals to join and participate in a particular organizational crowd-engagement endeavor, has an impact on the sort of crowd eventually constructed. While an open call may bring larger overall crowds, a more focused call, (say to a certain groups of experts) perhaps increases the probability of missing the best solution/idea, due to smaller overall crowd size.

Third, measures that capture the size and quality of crowds are also paramount given the known differences in crowd-size that exist [42], the known presence of malicious members in some crowds [65], and the relative specialization of some crowd intermediaries [16].

Taken altogether, these are some of the known measures that have been illustrated in research and practice in respect to differentiating crowds from one another, and we expect that more fine-grained measures can similarly emerge.

## 4.2 Measuring Crowd Capability

As already illustrated, Crowd Capability also presents a multitude of options for researchers and practitioners. To manage this large variation, we suggest operationalizing Crowd Capability by attributes that we can control for in our models.

First, IT-structures vary along the Collaborative and Episodic dimensions, as well as by the type of IT carrying said IT-structure, for example, web, mobile, SMS, intermediaries, etc. As others have illustrated [47, 62] the use of categorical variables can be useful for these purposes.

Second, categorical variables may also be best to capture the process dimension of the Crowd Capability construct, as illustrated by [16]. In addition, count variables that incorporate the number of steps that an organization takes to arrive at Crowd Capital can also capture the process mechanisms.

Third, for measurement of the content dimension, pre-existing content taxonomies may already be available from data providers. For example, taxonomies from tournament crowdsourcing intermediaries such as eYeka and Innocentive can be employed because these crowdsourcing intermediaries categorize their contests based on its salient attributes. This is perhaps the first way in which we can approach the operationalization of the content dimension of the Crowd Capability. In some cases, it may make sense to hold the content variable constant in the model so that the impact of changes in IT-structures or the target crowd, on the Crowd Capital generated, can be gauged accurately.

## 4. 3. Measuring Crowd Capital

As already illustrated, the unique organizational-level resource generated through crowd-engagement—Crowd Capital—is the benefit that organizations seek from IT-mediated crowds. When problem solving, creating ideas, or micro-tasking, organizations can build a multitude of specific forms of Crowd Capital from these potentials.

Researchers and managers operating in different industries and seeking different resources from a crowd will naturally need to develop the operationalization that makes most sense for them. However, in-line with [47, 62] among others, we posit that crowd participation/contribution count variables, count variables of the number of problems solved, and count variables of text accumulated, all qualify as meaningful measures of the Crowd Capital generated.

Now that our review of some crowd science operationalizations is complete, in the ensuing section we investigate some different research methods for an empirical crowd science.

## 5. Methods for Crowd Science

The ideal experiment to ascertain the benefits of crowd-engagement involves randomly selecting organizations seeking specific and similar resources from IT-mediated crowds, and observing how they do in terms of Crowd Capital generation afterwards. This is tantamount to the 'treatment' group [52]. Naturally, to have a baseline comparison point, we need to have a group of organizations—also with similar specific knowledge needs—that do not receive the treatment of crowd-engagement. This control group provides us with the counterfactual scenario.

Furthermore, within this experiment, we can incorporate several second-order modifications to ascertain the efficacy of certain crowds and capabilities for producing specific Crowd Capital needs. In other words, within the group that receives the treatment of crowd-engagement, we could change *one* of many dimensions such as Episodic vs. Collaborative IT-structure, type of crowd targeted, and so on, to arrive at the specific impact of *that* element of the crowd-engagement apparatus on Crowd Capital generated.

As one can imagine, this is a difficult experiment to conduct. First, we need to assemble a large sample of *similar* organizations who need the same Crowd Capital and convince them to either take the pill or the placebo. In reality, organizations have idiosyncratic needs and are free to follow whatever strategy they want to follow. Further, randomly modifying the crowd-engagement apparatus as described above would not only be difficult but also costly, if the modification is not a good fit for the organization's Crowd Capital strategy.

### 5.1 Toward the Ideal Experiment: Other Methods

Despite the growth and ubiquity of crowd-engagement, there is a striking absence of research toward developing a formal theory of crowd science. It seems that a problem that all scholars and practitioners of crowd science face in this respect, is that there is lack of data to conduct formal econometric analyses that strive to mimic the experimental ideal. For instance, in investigations of Crowd Capabilities that are web or intermediary-based, most datasets in today's research are collected from one organization or one intermediary only. If researchers are interested in online tournaments, they look solely to intermediaries like eYeka, Innocentive, or TopCoder, one at a time. If they are researching crowdfunding, they look only at Indiegogo and Kickstarter.

However, any formal theory of crowd science, if it is to pass the external validity test [42, 43, 48] has to speak to predictability across platforms, motivations, goals, institutions, and cultures involved in crowd-engagement endeavors. In this day and age, researchers are happy to find *one* platform as a data source, and so how are we to find datasets across platforms, goals, institutions, and cultures? The challenge becomes even graver when we move out of the crowd-intermediary sphere, and try to investigate the efficacy of organizations using their own in-house IT structures to engage a crowd. The Crowd Capabilities and the resources sought, both vary considerably from one organization to another, making the location of comparable treatment and control groups even tougher.

One of the ways to address this limitation is by introducing the advanced quantitative methods of meta-analysis, used widely in other social sciences [53, 34] to build datasets that allow us to perform some research-synthesizing and paradigm-developing research in the field. A meta-analysis harvests results from the extant literature in a field and aids research-synthesizing and theory advancing analyses. Next, we discuss this methodology in more detail.

### 5.2 Meta-Analysis

Given the theoretical adolescence, yet burgeoning nature of crowd science research, a meta-analysis, which evaluates the balance of evidence, is befitting [55, 56]. Starting with Hedges and Olkin (1985), homogeneity analyses are appropriate to synthesize evidence from extant research and arrive at cumulative verdicts on, for instance, the impact of targeted vs. open crowds for scientific problem solving (i.e. Citizen Science). Today, in addition to the traditional homogeneity analyses, more advanced meta-analytic regression analyses (MARA) [57] are also available to us. Thus, we are in a position to use meta-analysis more as a theory building technique rather than a vote counting tool [58].

In putting the currently available and increasingly developing tools of meta-analyses to use for our purposes, building a data set from the extant literature on the use and efficacy of crowd-engagement across platforms, cultures, and institutional contexts, is an area that merits future research work. A data set so collected, could readily harvest evidence on the aforementioned attributes of crowd science from hundreds of papers on the subject, and contain the most exhaustive data across capabilities and crowds, allowing us to take steps toward a more formal theory of crowd science. Further, we need not stop here, meta-analyses also allows us to combine the data collected from the extant published research with secondary data on economics, culture, and institutions, to perhaps make theoretical advances that would otherwise elude us.

### 5.3 Natural Experiments & Design Science

In addition to meta-analytic approaches, natural experiments also arise that allow researchers to gauge the efficacy of crowd-engagement. For instance, holding Crowd Capability constant, researchers have recently shown that crowd-engagement results in a larger pool of suggestions, but that organizations are still likely to pay more attention to solutions and knowledge contributions more familiar to them, hence defeating the purpose of crowd-engagement [17]. Work like this sheds light on efforts where crowd-engagement seems to have failed.

Furthermore, we reason that large organizations in retail or professional services can provide a particularly fertile setting for natural experiments. For instance, in a recent randomized trial, researchers found that in stores similar in terms of size, geography, and inventory, the use of an engaging marketing device led to a more than 25% increase in customer acquisition and sales, than in stores where the engaging marketing device was not installed [59]. Since, the device was installed on one type of inventory and the stores were all owned by a recognized national outlet, the randomized installation provides causal and assertive evidence of customer engagement. Established econometric techniques such as difference-in-differences methods [52] can readily be employed with such data, and we feel that similar studies could be undertaken specifically about IT-mediated crowd interventions.

Further, similar experiments can also be employed in large organizations with offices in multiple cities. From a set of regional offices, similar in most attributes such as size, function etc., some can be randomly chosen for the crowd-engagement treatment (such as implementing a local wiki). Similarly, difference-in-differences methods could compare the impact of the treatment with the counterfactual to reveal the benefit/costs of crowd-engagement for this organization.

And finally, we feel that design science [60, 61] is also a useful methodological option to investigate crowd science because much of the extant crowd-engagement research, particularly from the HCI or Computer Science disciplines, is already investigated though engineered artifacts aimed at various aspects of crowd-engagement [66]. More needs to be done to cross-apply these artifacts to other settings, while also cross-pollenating the research results that already exist in this vein, perhaps through a specific meta-analysis focused on artifacts engineered for crowd–engagement.

## 6. Limitations & Discussion

As with any other study, ours is not without limitations. First, the operationalizations, datasets, model, and methods that we put forth in this paper, though extensive, are surely not exhaustive. We feel that our recommendations will only get better once more current crowd-engagement researchers join our attempts at a unifying crowd science project. In short, we feel that the outline for empirics provided in this paper is just the beginning. Indeed, many intermediary-specific empirical works in the areas of Crowdsourcing (M-Turk, Wikipedia, and Innocentive), Citizen Science (Zooniverse), and Crowdfunding (Kickstarter, Indiegogo) for example, are already completed or are in progress. However, the work has been scattered and focused on specific parts of the crowd science apparatus without an anchor to a larger picture. The purpose here has been to lay the foundation for future work to be more purposive for the overall field, while developing datasets and natural experiments that will, piece by piece, allow us to add testability and predictability to crowd science.

And finally, there is the question of whether the development of a crowd science is worthwhile at all? And if so, is it unique?

In respect to the first question we feel that pursuing a crowd science is indeed worthwhile, for the simple fact that crowd science seems to span across disciplinary boundaries tackling similar phenomena. And these crowd science sub-disciplines, whether it is crowdsourcing, citizen science or crowdfunding, have already illustrated unique new potentials for organizational resource development that merit further investigation.

In respect to the second question, we feel that a crowd science is indeed unique in the study of socio-technical systems research, due to the somewhat breathtaking new potentials and outcomes that the disparate phenomena have brought to the public and private sector. We feel that moving toward a true crowd science can help researchers and practitioners alike in understanding the unprecedented on-demand scale of human participation, the unprecedented on-demand speed and aggregation of human effort, the unprecedented on-demand access to human knowledge, the new outcomes, and the new configurations of socio-technical systems that we routinely see in crowd science [67].

## 7. Conclusion

We started this paper with the implicit research question: what are the data, models, measures and methods that would allow a crowd science to develop in the future? In other words, what empirical apparatus would we need as IT-mediated crowd scientists to conduct research that causally asserts the benefits/costs of crowd-engagement across organizations, industries, platforms, geographies, cultures, and research silos?

In addressing this question, we anchored ourselves in TCC for sake of generalizability—mandatory in a field so diverse in methods and goals as crowd science. We first illustrated the measurements of all the pertinent constructs that lead to Crowd Capital generation. Second, we outlined the empirical challenges in causal identification of the costs and benefits of crowd-engagement. We then introduced datasets and methods that would allow IT-mediated crowd scientists to get closer to the experimental ideal, where one is able to assert the impact of one specific modification in the apparatus, on the outcome with certitude.

Our work makes several contributions to the literature on crowd-engagement. In particular, this is the first work to outline the research landscape and the empirical apparatus necessary for forming a generalizable and falsifiable crowd science. Similarly, our work also opens up numerous avenues for future research. For instance, building upon and augmenting the apparatus proposed here and conducting some of the empirical research clarified here, as natural next steps for crowd science researchers.